\documentclass[british,aps, pra]{revtex4}
\usepackage[T1]{fontenc}
\usepackage[latin9]{inputenc}
\usepackage{graphicx}
\usepackage{amssymb}

\makeatletter

\providecommand{\tabularnewline}{\\}

\@ifundefined{textcolor}{}
{%
 \definecolor{BLACK}{gray}{0}
 \definecolor{WHITE}{gray}{1}
 \definecolor{RED}{rgb}{1,0,0}
 \definecolor{GREEN}{rgb}{0,1,0}
 \definecolor{BLUE}{rgb}{0,0,1}
 \definecolor{CYAN}{cmyk}{1,0,0,0}
 \definecolor{MAGENTA}{cmyk}{0,1,0,0}
 \definecolor{YELLOW}{cmyk}{0,0,1,0}
 }

\makeatother

\usepackage{babel}

\begin{document}

\title{Indirect Quantum Tomography of Quadratic Hamiltonians}

\author{Daniel Burgarth$^{1,2}$, Koji Maruyama$^{2}$ and Franco Nori$^{2,3}$ }

\affiliation{$^{1}$IMS and QOLS, Imperial College, London SW7 2PG, UK }

\affiliation{$^{2}$Advanced Science Institute, RIKEN, Wako-shi, Saitama 351-0198,
Japan}

\affiliation{$^{3}$Physics Department, University of Michigan, Ann Arbor, Michigan,
48109, USA}
\begin{abstract}
A number of many-body problems can be formulated using Hamiltonians
that are quadratic in the creation and annihilation operators. Here,
we show how such quadratic Hamiltonians can be efficiently estimated
indirectly, employing very few resources. We find that almost all
properties of the Hamiltonian are determined by its surface, and that
these properties can be measured even if the system can only be initialised
to a mixed state. Therefore our method can be applied to various physical
models, with important examples including coupled nano-mechanical
oscillators, hopping fermions in optical lattices, and transverse
Ising chains. 
\end{abstract}
\maketitle

\section{Introduction}

There has been considerable interest in the problem of Hamiltonian
identification through indirect probing , thereby developing various
quantum mechanical versions of classical system tomography or classical
`inverse scattering' problems~\cite{Gladwell2004}. For certain types
of interactions, it was found \cite{Burgarth2009,Burgarth2009a,Wiesniak2010,Franco2009,Ashhab2006,Oxtoby2009}
that only few resources are required to obtain an accurate model of
the system. Indirect Hamiltonian estimation is therefore an interesting
problem for both pragmatic purposes and fundamental insights. We are
interested in the following questions: How can we obtain precise information
about a Hamiltonian under restricted access? What can we learn about
the `inside' of a large system by only looking at a subsystem of it?
Under which conditions is such indirect probing possible? %
\begin{figure}
\centering{}\includegraphics{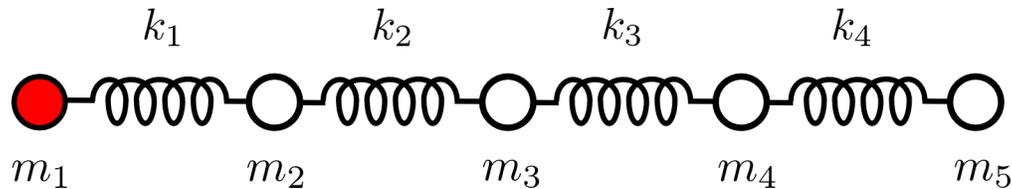}\caption{\label{fig:Diagram.-1}Indirect \emph{classical} system tomography
of a quadratic Hamiltonian. In this example, the spring constants
$k_{i}$ and masses $m_{i}$ of a chain of coupled harmonic oscillators
can be determined by monitoring the dynamics of a single particle
at the end (in red). See \cite{Gladwell2004} for details.}

\end{figure}

Recent studies have focused on this problem for cases of chains and
networks of spin-1/2 particles. The common question addressed can
be formulated as follows: Can we estimate all parameters, such as
coupling strengths and local fields, by accessing only one or a few
spins? It should be emphasised that even \emph{direct }Hamiltonian
estimation, or more generally, process tomography, is hard, because
the required number of measurements and the complexity of the post-processing
both scale exponentially with the system size. However, in realistic
situations, we usually have a priori knowledge based on the underlying
physics. It has been shown that such knowledge can be used to develop
compressed sensing protocols~\cite{Gross2009,Shabani2010}, which
greatly reduce the complexity of process tomography. Various works
on indirect Hamiltonian estimation have relied on similar assumptions;
namely, that the dynamics is restricted to a subspace of polynomial
dimension \cite{Burgarth2009,Burgarth2009a,Wiesniak2010}. In \cite{Burgarth2009},
the efficiency of the estimation in terms of the required time and
the number of measurements is discussed. An interesting example that
does not rely on a subspace was analysed by Di Franco et al. \cite{Franco2009}.
We will see here that this is a special case of the generic estimation
of quadratic Hamiltonians, which can be estimated efficiently due
to a simple description of their dynamics in the Heisenberg picture.
Di Franco also found that the estimation is quite robust against noise
\cite{Franco2009}. In \cite{Burgarth2009a} the 1D methods were generalised
to arbitrary graphs, and the possible elimination of degeneracies
was discussed. Also, Wiesniak and Markiewicz \cite{Wiesniak2010}
went beyond the simplest subspace in order to study quasi-1D systems.
Table \ref{tab:Overview-of-indirect} summarises the results obtained
so far in terms of the settings and assumptions considered.%
\begin{table}
\begin{centering}
\begin{tabular}{|c|c|c|c|c|}
\hline 
Interaction type & Needs preparation & Geometry & Obtain & Reference\tabularnewline
\hline
\hline 
$XX+YY+\Delta ZZ$ & specific state & 1D & couplings & \cite{Burgarth2009}\tabularnewline
\hline 
$\begin{array}{c}
(1+\gamma)XX+(1-\gamma)YY\\
\gamma\neq1,-1\end{array}$ & no & 1D & couplings & \cite{Franco2009}\tabularnewline
\hline 
$XX+YY+\Delta ZZ+Z$ & specific state & arbitrary & couplings and fields & \cite{Burgarth2009a}\tabularnewline
\hline 
$XX+YY+Z$ & specific state & quasi-1D & couplings, partial topology & \cite{Wiesniak2010}\tabularnewline
\hline 
$\begin{array}{c}
a^{\dagger}a+aa+h.c.\\
\mbox{(fermions or bosons)}\\
(1+\gamma)XX+(1-\gamma)YY+Z\end{array}$ & arbitrary state & arbitrary & couplings, fields, anisotropies & this paper\tabularnewline
\hline
\end{tabular}
\par\end{centering}

\caption{\label{tab:Overview-of-indirect}Overview of indirect Hamiltonian
estimation schemes. The interaction types represent the Pauli matrices
involved in the spin-spin coupling, e.g., $XX+YY+\Delta ZZ+Z$ stands
for a Hamiltonian of the form $\sum_{n,m}A_{nm}\left(XX+YY+\Delta ZZ\right)_{n,m}+\sum_{n}B_{n}Z_{n}.$}

\end{table}
 However, the analysis of physically important cases, such as the
transverse Ising model and the XY model with a magnetic field, have
remained open. Solutions to both cases will be presented in this work. 

Our main goal in this paper is to develop a method to perform indirect
quantum tomography for many-body systems of identical particles. Even
though the method is analogous to the spin case, the Hamiltonians
considered here have a higher number of parameters, and it is surprising
that they can still be estimated in a similar fashion. The class of
Hamiltonians we study here is those of quadratic form in bosonic or
fermionic operators. There has been tremendous progress in experiments
of quantum random walks~\cite{Karski2009}, optical lattices~\cite{Garcia-Ripoll2003},
coupled cavities~\cite{Plenioa}, nano-mechanical oscillators~\cite{Connell2010},
etc., which can be modelled by such quadratic Hamiltonians. Thus,
the indirect estimation scheme we present here will be of use in reducing
the necessary resources for modelling such systems. For the case of
bosons, it is the most direct translation of the work by Gladwell~\cite{Gladwell2004}
to the quantum case. Gladwell studied how the spring constants and
masses of coupled classical harmonic oscillator chains can be estimated
by looking at the movement of only one particle (see also Fig. \ref{fig:Diagram.-1}).
Furthermore, our estimation protocol gives a natural generalisation
of the problems on spin chains, since quadratic Hamiltonians of fermions
also describe a certain class of spin systems, such as the transverse
Ising model. 

\begin{figure}
\centering{}\includegraphics{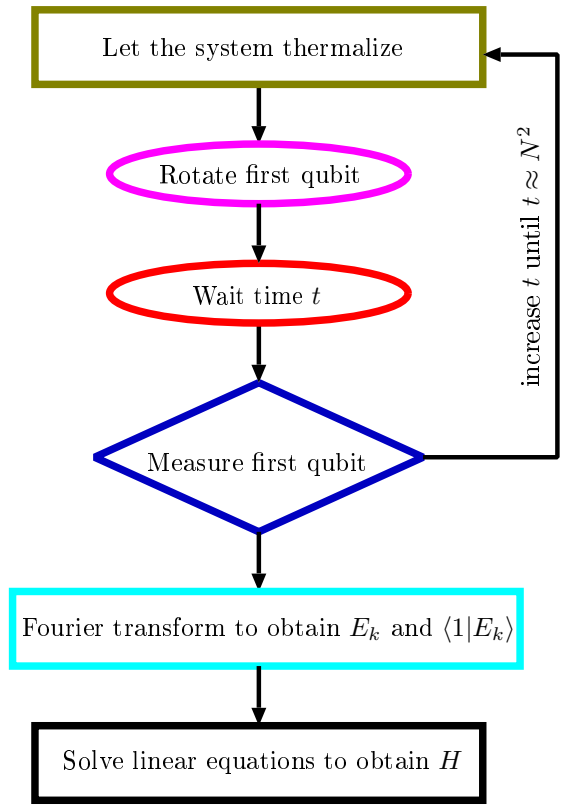}\caption{\label{fig:Diagram.}Schematic overview of our estimation scheme for
a chain of length $N.$ The main requirements are (i) relatively fast
single-qubit rotations and measurements on the first site, and (ii)
a decoherence time of at least $N^{2}.$ }

\end{figure}

The main method of indirect estimation is summarised as follows. First,
the system is initialised to an arbitrary but fixed state. This can
be, for example, even a thermal state, and can occur on slow time-scales
via relaxation. Then, some simple single-particle properties are initialised
and measured at a later time. Finally, the accumulated data is Fourier
transformed, and the parameters are extracted through a set of linear
equations. This simple method is outlined in Fig. \ref{fig:Diagram.}
for the 1D case. %
 The procedure is analogous to an `inverse scattering' problem because
the perturbation introduced in one edge of the sample (e.g., rotation
of the first qubit) propagates through the sample, `scattering' with
the inner structure of the Hamiltonian, and then this information
encodes the structure of the system. While it is obvious that this
procedure provides \emph{some }information on the system, the surprising
result here is that \emph{all }information can be uniquely identified.
Since we obtain information on anisotropies as well, not only the
coupling strengths but also the type of interaction can be determined
by our method.

The paper is structured as follows. First, we introduce the necessary
notation for quadratic Hamiltonians and some techniques for their
diagonalisations. Since these are well established methods, we will
only discuss them as a `recipe' for the estimation procedure. We then
discuss the simplest case of estimation, namely when the system is
a chain of hopping particles, and later generalise it to arbitrary
graphs. Finally we discuss how the results apply to 1D chains of spins
and conclude.

\section{Notation and Diagonalisation}

The most general quadratic Hamiltonian of $N$ indistinguishable particles
is written as\begin{equation}
H=\sum_{n,m=1}^{N}A_{nm}a_{n}^{\dagger}a_{m}+\frac{1}{2}\sum_{n,m=1}^{N}\left(B_{nm}a_{n}^{\dagger}a_{m}^{\dagger}+B_{nm}^{*}a_{m}a_{n}\right),\label{eq:ham_original}\end{equation}
where the $a_{n}^{\dagger}$ and $a_{m}$ are creation and annihilation
operators and $A,B$ are matrices describing the parameters we would
like to estimate. For $H$ to be Hermitian we must have $A=A^{\dagger}$
and $B^{T}=-\epsilon B,$ where we introduced the parameter $\epsilon=1$
for fermions and $\epsilon=-1$ for bosons. We will mostly follow
the notation of \cite{Blaizot1986}, though we shall write all vectors
in Dirac notation. At first, we put all Hamiltonian parameters into
the Hermitian $2N\times2N$ matrix

\begin{equation}
M\equiv\left(\begin{array}{cc}
A & B\\
-\epsilon B^{*} & -\epsilon A^{*}\end{array}\right),\label{eq:matrix_m}\end{equation}
and introduce the column vector operator \[
\alpha\equiv\left(\begin{array}{c}
a_{1}\\
\vdots\\
a_{N}\\
a_{1}^{\dagger}\\
\vdots\\
a_{N}^{\dagger}\end{array}\right),\]
so that Eq.~(\ref{eq:ham_original}) can be expressed up to a constant
as \[
H=(1/2)\alpha^{\dagger}M\alpha.\]
 Throughout this paper, we make two technical assumptions: first,
all the \emph{phases }of $M_{ij}$ are assumed to be known. Although
some phases might be easy to determine, this requires complicated
studies of gauge invariance that do not seem to be worthwhile, as
in many practical cases all elements are real and positive. Second,
for the bosonic case we assume that the matrix $M$ is \emph{positive
definite}. Again, in principle this can be generalised, but this way
we avoid technical difficulties of symplectic transformations \cite{Blaizot1986}. 

As in \cite{Burgarth2009a} the efficiency of our method depends on
how many entries of $M$ are a priori known to be zero; that is, on
knowledge of the coupling graph. If such knowledge is not available,
we have to perform measurements on all but one of the qubits. If the
graph is known to be highly sparse (for instance a chain) we only
need to access a single qubit. But before going into the details of
Hamiltonian identification, let us briefly review the diagonalisation
of the Hamiltonian Eq.~(\ref{eq:ham_original}) and thus the dynamics,
introducing some notations. For more detailed descriptions on the
diagonalisation procedure, see, e.g., \cite{Blaizot1986}. 

For quadratic Hamiltonians of the form in Eq.~(\ref{eq:ham_original}),
there exist quasi-particle creation and annihilation operators $b_{k}^{\dagger}$
and $b_{\ell}$, with which the Hamiltonian can be represented by
the simple form of non-interacting modes, \begin{equation}
H=\sum_{k}E_{k}b_{k}^{\dagger}b_{k}+\mbox{const.}\label{eq:ham_diag}\end{equation}
For this reason, quadratic Hamiltonians are also referred to as `quasi-free'
interactions. We need to know the transformation $T$ that maps the
operators $a$ and $a^{\dagger}$ for particles to $b$ and $b^{\dagger}$
for quasi-particles, i.e., $\beta=T\alpha,$ where $\beta$ is defined
by \[
\beta\equiv\left(\begin{array}{c}
b_{1}\\
\vdots\\
b_{N}\\
b_{1}^{\dagger}\\
\vdots\\
b_{N}^{\dagger}\end{array}\right).\]
 In order to ensure the canonical commutation relations for the operators
$b_{k}$ and $b_{k}^{\dagger}$, $T$ must satisfy $T^{-1}=\eta T^{\dagger}\eta$,
where \[
\eta=\left(\begin{array}{cc}
\openone & 0\\
0 & \epsilon\openone\end{array}\right).\]
 The Hamiltonian is now written as \[
H=\frac{1}{2}\beta^{\dagger}\eta(T\eta MT^{-1})\beta+\mbox{const.}\]
It can then be shown that $\eta M$ is diagonalised by $T$ as \[
T\eta MT^{-1}=\left(\begin{array}{cc}
E & 0\\
0 & -E\end{array}\right),\]
where $E=\mbox{diag}\left\{ E_{1},...,E_{N}\right\} $, to have the
desired form of Eq.~(\ref{eq:ham_diag}). Note that the energy eigenvalues
appear in pairs of positive $E_{k}$ and negative $E_{k+N}\equiv-E_{k}$
values $(k=1,...,N).$ 

The matrix $T$ consists of the \emph{right eigenvectors} $|E_{k}\rangle$
of $\eta M$ as\[
T\equiv\eta\left(\begin{array}{c}
\langle E_{1}|\\
\vdots\\
\langle E_{2N}|\end{array}\right)\eta,\]
and the inverse of $T$ is given by\[
T^{-1}=\left(|E_{1}\rangle\cdots|E_{2N}\rangle\right).\]
For bosons, the matrix $\eta M$ is not Hermitian and the distinction
between right and left eigenvectors is necessary. This gives rise
to a few further peculiarities, such as the modified normalisation
and completeness relationship (see below). Fortunately, in this work
we are solving an \emph{inverse problem, }and do not have to discuss
how to find these vectors and how numerically stable the corresponding
algorithms are.

It is worth pointing out that the $|E_{k}\rangle$ are not representing
physical states but just introduced here as a part of solving the
Heisenberg equation of motion for the creation and annihilation operators.
The completeness relationship is given by\begin{equation}
\sum_{k=1}^{N}|E_{k}\rangle\langle E_{k}|\eta+\epsilon|E_{k+N}\rangle\langle E_{k+N}|\eta=\openone_{2N\times2N},\label{eq:complete}\end{equation}
and the vectors $|E_{k}\rangle$ are chosen to fulfil the normalisation
relationship\begin{eqnarray}
\langle E_{k}|\eta|E_{k'}\rangle & = & \eta_{kk'}.\label{eq:norm}\end{eqnarray}

For convenience, let us also introduce vectors $|n\rangle$ as the
canonical basis vectors: \[
|n\rangle\equiv\left(\begin{array}{c}
0\\
\vdots\\
0\\
1\\
0\\
\vdots\\
0\end{array}\right)\,\leftarrow n\mbox{-th row. }\]
Due to the structure of the matrix $M,$ the upper and lower eigenvectors
of $\eta M$ are related as \[
\langle n|E_{k\oplus N}\rangle=\langle n\oplus N|E_{k}\rangle^{*},\]
where $\oplus$ is the addition modulo $2N.$ The dynamics of the
original operators $\alpha$ can be found from $\beta_{n}(t)=e^{-iE_{m}t}\beta_{n}(0)$
as \begin{equation}
\alpha_{n}(t)=\sum_{m,k}s(m,k)e^{-iE_{k}t}T_{nk}^{-1}\left(T_{km}^{-1}\right)^{\dagger}\alpha_{m}(0),\label{eq:time_evltn_alpha}\end{equation}
where we have introduced a sign function $s(m,k)$ through\begin{eqnarray*}
s(m,k) & = & 1\qquad(m=1,\ldots N;k=1,\ldots N)\\
s(m,k) & = & \epsilon\qquad(m=1,\ldots N;k=N+1,\ldots2N)\\
s(m,k) & = & \epsilon\qquad(m=N+1,\ldots2N;k=1,\ldots N)\\
s(m,k) & = & 1\qquad(m=N+1,\ldots2N;k=N+1,\ldots2N).\end{eqnarray*}

\section{\label{sec:Estimation-of-chains}Estimation of chains}

\subsection{Experimental requirements\label{sub:Experimental-requirements}}

Let us first consider a 1D chain of interacting particles, meaning
that $A$ and $B$ are tridiagonal. Also assume that we can initialise
the chain in a fixed state $\rho_{0}.$ This state could, for instance,
be a thermal state, but we do not require to have the exact form of
$\rho_{0}$: we just have to be able to repeatedly initialise the
chain to the same state $\rho_{0}$. As before~\cite{Burgarth2009,Franco2009},
we perform initialisations followed by measurements at the first site.
The quantity we need to measure at the first site is $a_{1}\pm a_{1}^{\dagger}$
for different times up to $N^{2}$ \cite{Burgarth2009}. In order
to eradicate the dependence on the initial state $\rho_{0},$ we measure
two sequences of $\langle a_{1}(t)\rangle$ after preparing the first
site to give two different initial values, i.e., $\langle a_{1}(0)\rangle=c_{1}$
and $\langle a_{1}(0)\rangle=c_{2}$. Using $\left\langle a_{n}\right\rangle =\overline{\left\langle a_{n}^{\dagger}\right\rangle },$
and thus $\langle a_{1}^{\dagger}(0)\rangle=c_{i}^{*}$, and subtracting
the measurement results, we obtain a quantity that only depends on
$\Delta c\equiv c_{1}-c_{2}.$ It is given through Eq.~(\ref{eq:time_evltn_alpha})
by\begin{eqnarray}
\left\langle a_{1}(t)\right\rangle _{c_{1}}-\langle a_{1}(t)\rangle_{c_{2}} & = & \left[\sum_{m,k=1}^{2N}s(m,k)e^{-iE_{k}t}T_{1k}^{-1}\left(T_{km}^{-1}\right)^{\dagger}\left\langle \alpha_{m}(0)\right\rangle \right]_{1}-\left[\sum_{m,k=1}^{2N}...\right]_{2}\nonumber \\
 & = & \Delta c\sum_{k=1}^{2N}s(1,k)e^{-iE_{k}t}|T_{1k}^{-1}|^{2}+\Delta c^{*}\sum_{k=1}^{2N}s(N+1,k)e^{-iE_{k}t}T_{1k}^{-1}\left(T^{-1}\right)_{k,N+1}^{\dagger}.\label{eq:meas1}\end{eqnarray}
This initialisation can be performed by a von Neumann measurement,
or, as long as the reduced density matrix at site one is not maximally
mixed, by applying different single qubit rotations (in some experiments
von Neumann measurements are hard). As we see in Eq.~(\ref{eq:meas1}),
the dependence on the initial state is completely removed. This is
thanks to the absence of the interactions between particles: they
(almost) do not see each other, so the information on the `injected'
particle can be extracted by subtracting the influence from others. 

For the spin chain case the eigenfrequencies are non-degenerate and
$T_{1k}^{-1}=\langle1|E_{k}\rangle\neq0\:(\forall k)$ \cite{Burgarth2009,Burgarth2009a}.
For the present case of quadratic Hamiltonians we were unable to prove
this, but could confirm it numerically. Hence, a Fourier analysis
provides us with the frequencies $E_{k}$ and the amplitudes $\Delta c\,|T_{1k}^{-1}|^{2}+\Delta c^{*}\epsilon\, T_{1k}^{-1}\left(T^{-1}\right)_{k,N+1}^{\dagger}.$
Summing these amplitudes gives the value of $\Delta c:$\begin{eqnarray*}
\sum_{k=1}^{N}\left(\Delta c\,|T_{1k}^{-1}|^{2}+\Delta c^{*}\epsilon\, T_{1k}^{-1}\left(T^{-1}\right)_{k,N+1}^{\dagger}\right)+\epsilon\sum_{k=N+1}^{2N}\left(\Delta c\,|T_{1k}^{-1}|^{2}+\Delta c^{*}\epsilon\, T_{1k}^{-1}\left(T^{-1}\right)_{k,N+1}^{\dagger}\right) & =\\
\Delta c\,\langle1|1\rangle+\Delta c^{*}\epsilon\,\langle1|N+1\rangle & = & \Delta c,\end{eqnarray*}
where we used the completeness relationship Eq.~(\ref{eq:complete}).
Equation~(\ref{eq:meas1}) still contains mixtures of the coefficients
$|T_{k1}^{-1}|$ and $T_{1k}^{-1}\left(T^{-1}\right)_{k,N+1}^{\dagger}.$
We can separate them by measuring another pair of initialisations
$c'_{i}:$ as long as $\Delta c'\neq r\Delta c\;(r\in\mathbb{R}),$
we can solve the linear equation for $|T_{k1}^{-1}|.$ Without loss
of generality, we choose $|\langle1|E_{k}\rangle|=\left|T_{1k}^{-1}\right|=T_{1k}^{-1}=\langle1|E_{k}\rangle\:(\forall k)$
by arranging the global phase of each eigenstate $|E_{k}\rangle$.
In conclusion, a few random rotations or initialisations of the first
qubit, followed by measurements, provide us with $E_{k}$ and $\langle1|E_{k}\rangle.$

Let us now describe how to obtain the parameters of $M$ from this
observed data. We have to distinguish between the generic case where
the off-diagonal couplings $A_{n,n+1}$ and $B_{n,n+1}$ are distinct,
and the special case where they are equal.

\subsection{Different off-diagonal couplings}

\begin{figure}
\centering{}\includegraphics[width=11cm]{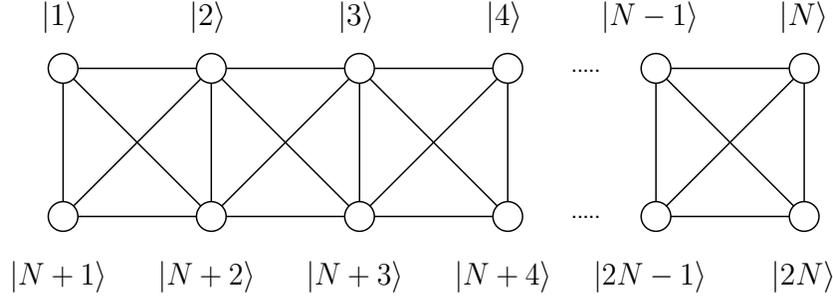}\caption{\label{fig:couplingmatrix}The matrix $M$ as a coupling graph on
the nodes $|n\rangle.$}

\end{figure}

As we have seen above, what we diagonalised is the $2N\times2N$ matrix
$\eta M$, so it is helpful to regard $M$ as a representation of
a graph consisting of $2N$ nodes (see Fig.~\ref{fig:couplingmatrix}).
Its off-diagonal entries correspond to the coupling strengths between
nodes, whereas the diagonal elements represent the intensity of the
`field' at each node. %
 We can then start with a recursive algorithm similar to \cite{Gladwell2004,Burgarth2009,Burgarth2009a}
by applying $M$ to the localised states at sites 1 and $N+1;$ \begin{eqnarray*}
M|1\rangle & = & A_{11}|1\rangle+A_{21}|2\rangle-\epsilon B_{11}^{*}|N+1\rangle-\epsilon B_{21}^{*}|N+2\rangle\\
M|N+1\rangle & = & B_{11}|1\rangle+B_{21}|2\rangle-\epsilon A_{11}^{*}|N+1\rangle-\epsilon A_{21}^{*}|N+2\rangle.\end{eqnarray*}
Using these equations for $\langle1|\eta M|E_{k}\rangle$ and $\langle N+1|\eta M|E_{k}\rangle$
we arrive at

\begin{eqnarray*}
E_{k}\langle1|E_{k}\rangle & = & \langle1|\eta M|E_{k}\rangle=\langle1|M|E_{k}\rangle=A_{11}\langle1|E_{k}\rangle+A_{21}^{*}\langle2|E_{k}\rangle-\epsilon B_{11}\langle N+1|E_{k}\rangle-\epsilon B_{21}\langle N+2|E_{k}\rangle\\
\epsilon E_{k}\langle N+1|E_{k}\rangle & = & \epsilon\langle N+1|\eta M|E_{k}\rangle=\langle N+1|M|E_{k}\rangle=B_{11}^{*}\langle1|E_{k}\rangle+B_{21}^{*}\langle2|E_{k}\rangle-\epsilon A_{11}\langle N+1|E_{k}\rangle-\epsilon A_{21}\langle N+2|E_{k}\rangle.\end{eqnarray*}
Note that $B_{11}=0$ for fermions. Among these parameters, we can
immediately obtain $A_{11}$and $B_{11}$ from the quantities that
are estimated by measuring $\langle a_{1}(t)\rangle$ as \begin{eqnarray*}
A_{11} & = & \langle1|M|1\rangle=\sum_{k=1}^{N}\left[E_{k}\left|\langle1|E_{k}\rangle\right|^{2}+\epsilon E_{k+N}\left|\langle1|E_{k+N}\rangle\right|^{2}\right],\\
B_{11}=\langle1|\eta M|N+1\rangle & = & -\sum_{k=1}^{N}\left[E_{k}\langle1|E_{k}\rangle\langle E_{k}|N+1\rangle+\epsilon E_{k+N}\langle1|E_{k+N}\rangle\langle E_{k+N}|N+1\rangle\right].\end{eqnarray*}
Therefore we can collect all the known terms of the above equations
on the left-hand-side as\begin{eqnarray}
\mbox{known} & = & A_{21}^{*}\langle2|E_{k}\rangle-\epsilon B_{21}\langle N+2|E_{k}\rangle\label{eq:a2ek}\\
\mbox{known} & = & B_{21}^{*}\langle2|E_{k}\rangle-\epsilon A_{21}\langle N+2|E_{k}\rangle.\label{eq:b2ek}\end{eqnarray}
Taking linear combinations, we obtain\begin{eqnarray*}
\mbox{known} & = & \left(A_{21}^{*}-\frac{\left|B_{21}\right|^{2}}{A_{21}}\right)\langle2|E_{k}\rangle\\
\mbox{known} & = & \left(B_{21}-\frac{\left|A_{21}\right|^{2}}{B_{21}^{*}}\right)\langle N+2|E_{k}\rangle.\end{eqnarray*}
It may appear as if the right-hand-side contains too many unknowns
to solve these equations. However, similar to the original work by
Gladwell \cite{Gladwell2004} and the spin chain case, \cite{Burgarth2009},
we can use the normalisation of the eigenstates. In this case, it
is given by Eq.~(\ref{eq:norm}). Summing up the mod squares of the
above equations, the dependencies on $\langle2|E_{k}\rangle$ and
$\langle N+2|E_{k}\rangle$ vanishes and we can infer the absolute
value of each coefficient, i.e., $\frac{1}{|A_{12}|}\left(\left|A_{21}\right|^{2}-\left|B_{21}\right|^{2}\right)$
and $\frac{1}{|B_{12}|}\left(\left|A_{21}\right|^{2}-\left|B_{21}\right|^{2}\right)$.
If $|A_{21}|\neq|B_{21}|,$ we obtain $|B_{21}/A_{21}|\equiv g$ by
dividing the above two equations and then through $|A_{21}|(1-g^{2}),$
both $|A_{21}|$ and $|B_{21}|$ are obtained. Because the phases
of $A$ and $B$ are known, then we learn $A_{21}$ and $B_{21}$.
We can then express a similar set of equations for the next site $\langle2|\eta M|E_{k}\rangle$
and $\langle N+2|\eta M|E_{k}\rangle.$ By induction, we then see
that all matrix elements of $A$ and $B$ can be obtained, as desired.
It is worth pointing out that the scheme works even though the graph
in Fig.~\ref{fig:couplingmatrix} is not infecting~\cite{Burgarth2009a}.

We will now look at the cases with equal off-diagonal couplings in
more detail, because such physical systems are often encountered,
e.g., transverse Ising for fermions, coupled harmonic oscillators
for bosons.

\subsection{Equal off-diagonal couplings}

When $|A_{n+1,n}|=|B_{n+1,n}|,$ the above method fails. This is the
case for interacting harmonic oscillators without the rotating wave
approximation~\cite{Eisert2004} and for quantum Ising models, and
therefore of interest in a number of practical situations. We focus
on the case where $A$ and $B$ are real, e.g., $A_{n,n+1}=B_{n,n+1}=-\epsilon B_{n+1,n}$.
The diagonal elements $A_{nn}$ and $B_{nn}$ are always different
if there is a transverse field (fermions) or if the masses are finite
(bosons), so it is reasonable to assume $A_{nn}\neq B_{nn}$ (the
Ising model without transverse field cannot be estimated using our
method because excitations do not propagate).

It is convenient to introduce\[
|n^{\pm}\rangle\equiv\frac{1}{\sqrt{2}}\left(|n\rangle\pm|n+N\rangle\right),\, n=1,\ldots,N.\]
 A simple calculation then shows that \[
M\eta|n^{\pm}\rangle=(1\pm\epsilon)A_{n,n-1}|n-1^{\mp}\rangle+(1\mp1)A_{n+1,n}|n+1^{\mp}\rangle+\left(A_{nn}\pm B_{nn}\right)|n^{\mp}\rangle,\, n=1,\ldots,N,\]
where we set $A_{01}=A_{N,(N+1)}=0.$ In some sense, this is similar
to a 1D chain case. As the elements $\langle1^{\pm}|E_{k}\rangle$
are already known, we learn $A_{11}$ and $B_{11}$ from\begin{eqnarray*}
\langle1^{\mp}|\eta M|1^{\pm}\rangle & = & A_{11}\pm\frac{1-\epsilon}{2}B_{11},\end{eqnarray*}
which can also be written in terms of the known variables by inserting
the completeness relation Eq. (\ref{eq:complete}). Using $E_{k}\langle1^{\pm}|E_{k}\rangle=\langle1^{\pm}|\eta M|E_{k}\rangle$
and normalisation we obtain $A_{21}$ and $|2^{+}\rangle.$ For bosons,
we then obtain $A_{22}-B_{22}=\langle2^{+}|\eta M|2^{-}\rangle$ through
the completeness relation, followed by $\langle2^{-}|E_{k}\rangle$
through \begin{eqnarray}
E_{k}\langle n^{+}|E_{k}\rangle=\langle n^{+}|\eta M|E_{k}\rangle= & (A_{nn}-B_{nn})\langle n^{-}|E_{k}\rangle,\label{eq:nminus}\end{eqnarray}
and finally \[
A_{22}+B_{22}=\langle2^{-}|\eta M|2^{+}\rangle,\]
by completeness again. On the other hand, for fermions,\[
E_{k}\langle2^{+}|E_{k}\rangle=\langle2^{+}|\eta M|E_{k}\rangle=2A_{21}\langle1^{-}|E_{k}\rangle+A_{22}\langle2^{-}|E_{k}\rangle,\]
and through normalisation we obtain $A_{22}$ and $\langle2^{-}|E_{k}\rangle.$
Knowing all parameters at site $2,$ we can then proceed through induction.
For the most general case, we can also allow for chains which sometimes
have equal off-diagonal couplings and sometimes different ones, by
alternating between the strategies described here and in the last
subsection.

\subsection{\label{sec:Estimation-with-Initialization;}Estimation of general
graphs}

We now briefly describe how the linear case is generalised to arbitrary
graphs. This is almost identical to \cite{Burgarth2009a}, so we will
not repeat the details. Similar to the spin case, in the general graph
setting, measurements on a single spin do not suffice: we need to
consider transport in the network. Depending on the network topology,
we choose a set $C$ of `infecting' \cite{Burgarth2009a} nodes, which
are the ones we will perform initialisations and measurements on.
For clarity, let us recall the definition of graph `infection'. Suppose
that a subset $C$ of nodes of the graph is `infected' with some property,
e.g., the flu. This property then spreads, infecting other nodes,
by the following rule: an infected node infects a `healthy' (uninfected)
neighbour if and only if it is its unique healthy neighbour. If eventually
all nodes are infected, the initial set $C$ is called `infecting'. 

Similar to the measurements described in Subsection~\ref{sub:Experimental-requirements},
initialising the site $m\in C$ and measuring the $\ell\mbox{th}$
node after some time, we can obtain\begin{eqnarray}
\left\langle a_{\ell}(t)\right\rangle  & = & \sum_{m,k}s(m,k)e^{-iE_{k}t}T_{\ell k}^{-1}\left(T_{km}^{-1}\right){}^{\dagger}\left\langle \alpha_{m}(0)\right\rangle \nonumber \\
 & = & \sum_{k}s(m,k)e^{-iE_{k}t}T_{\ell k}^{-1}\left(T_{km}^{-1}\right){}^{\dagger}\left\langle \alpha_{m}(0)\right\rangle +\sum_{k}s(m,k)e^{-iE_{k}t}T_{\ell k}^{-1}\left(T_{km}^{-1}\right){}^{\dagger}\left\langle \alpha_{N+m}(0)\right\rangle \nonumber \\
 & = & \sum_{k}s(m,k)e^{-iE_{k}t}T_{\ell k}^{-1}\left(T_{km}^{-1}\right){}^{\dagger}\left\langle a_{m}(0)\right\rangle +\sum_{k}s(m,k)e^{-iE_{i}t}T_{\ell k}^{-1}\left(T_{km}^{-1}\right){}^{\dagger}\left\langle a_{m}^{\dagger}(0)\right\rangle .\label{eq:3rec}\end{eqnarray}
Again, the dependence on the initial state $\rho_{0}$ may be removed
by subtracting data for different initial conditions on the sites
$m$ and $\ell$. Starting from some element in $C,$ say $m=\ell=1,$
we can get $T_{1k}^{-1}$ as described in Subsection~\ref{sub:Experimental-requirements}.
Then we initialise in $m$ and measure at a different site $\ell\in C,$
obtaining $T_{\ell k}^{-1}$ \emph{including its phase} from Eq.~(\ref{eq:3rec}).
Hence, all $T_{k\ell}$ with $k,\ell\in C$ can be learnt from simple
experiments on the set $C$. The $1D$ estimation and infection are
then used to infer the remaining parameters, as described in more
detail in \cite{Burgarth2009a}.

\section{Application to 1D spin chains}

Naturally, the above scheme can be applied directly to many cases
of Hamiltonian identification for systems of spin-1/2 particles. A
typical example is the $XY$ chain of spin-1/2 particles,\[
H=\sum_{n=1}^{N}c_{n,n+1}[(1+\gamma)S_{n}^{x}S_{n+1}^{x}+(1-\gamma)S_{n}^{y}S_{n+1}^{y}]+\sum_{n=1}^{N}b_{n}S_{n}^{z},\]
as it can be transformed into quasi-free fermions by means of the
Jordan-Wigner transformation. As it has been noted already in \cite{Franco2009},
the estimation for this model can be done without initialisation of
the entire chain. In the Jordan-Wigner picture this becomes very clear.
That is, after locally measuring an eigenstate of $X_{1}=a_{1}+a_{1}^{\dagger},$
thus making $\langle Z_{1}\rangle=0,$ the initial expectation values
of the $a_{n}$ and $a_{n}^{\dagger}$ $n>1$ are all zero. This is
because the Jordan-Wigner transformation of $a_{n}$ ($n>1)$, i.e.,
$a_{n}=\sigma_{n}^{+}\prod_{m<n}Z_{m},$ always contains $Z_{1}$
in the product. One might say that the local initialisation in the
spin picture corresponds to a global initialisation in the fermionic
picture. Combined with the weak dependence of local observables on
the initial condition that comes from the quasi-free interaction,
the state dependence of the measurements at the first site is completely
removed. Hence our scheme is a proper generalisation of \cite{Franco2009}
to include magnetic field and the transverse Ising case, which is
important in various physical systems, e.g., superconducting (flux)
qubits \cite{you2005}, NMR, etc. Such models have also attracted
attention in the context of indirect quantum control recently~\cite{Burgarth2010,Kay2010},
where our estimation scheme is crucial.

\section{Conclusions}

We found a simple and efficient method to identify the Hamiltonian
of a system of coupled bosons or fermions. While the methods are completely
analogous to the spin case \cite{Burgarth2009,Burgarth2009a}, it
is surprising that the higher number of parameters in the Hamiltonian
that arises from the non-conservation of excitations can still be
estimated using the same resources. Similarly to \cite{Franco2009},
we can deal with very weak system initialisation, such as thermal
states. Therefore our methods can drastically reduce the required
resources for system identification. Since we allow for site-dependent
anisotropies, not only the coupling strengths but also the type of
interaction is determined along the way. From the theory side, our
work once more confirmed a type of `area law' for estimation: looking
only at the surface of short-range interacting systems can completely
determine their Hamiltonian. It would be intriguing to see if this
has direct connections with area laws of entanglement \cite{Eisert2010}.
While the efficiency of our method relied on the quadratic form of
the Hamiltonian, we conjecture that even for models with true interaction
terms, e.g., quartic terms in the Hamiltonian, all system parameters
remain discoverable on the surface in principle.
\begin{acknowledgments}
We acknowledge support by the EPSRC grant EP/F043678/1, the JSPS Postdoctoral
Fellowship for North American and European Researchers (Short-Term)
(DB), the JSPS Kakenhi (C) No. 22540405 (KM), the National Security
Agency, Laboratory of Physical Sciences, Army Research Office, National
Science Foundation grant No. 0726909, JSPS-RFBR contract No. 09-02-92114,
MEXT Kakenhi on Quantum Cybernetics, and Funding Program for Innovative
R\&D on S\&T (FN).
\end{acknowledgments}

\end{document}